\documentclass[10pt,conference,compsocconf,letterpaper]{IEEEtran}

\usepackage{graphicx}
\graphicspath{{figs/}}
\DeclareGraphicsExtensions{.pdf,.jpeg,.png}

\usepackage{subfig}
\usepackage[flushleft]{threeparttable}

\usepackage{url}

\usepackage{setspace}
\usepackage{balance}

\usepackage{hyperref}
\usepackage{amsmath}

\usepackage{listings}
\usepackage[noend]{algpseudocode}
\usepackage{algorithm}
\usepackage{algpseudocode}

\algnewcommand\algorithmicswitch{\textbf{switch}}
\algnewcommand\algorithmiccase{\textbf{case}}
\algnewcommand\algorithmicassert{\texttt{assert}}
\algnewcommand\Assert[1]{\State \algorithmicassert(#1)}%
\algdef{SE}[SWITCH]{Switch}{EndSwitch}[1]{\algorithmicswitch\ #1\ \algorithmicdo}{\algorithmicend\ \algorithmicswitch}%
\algdef{SE}[CASE]{Case}{EndCase}[1]{\algorithmiccase\ #1}{\algorithmicend\ \algorithmiccase}%
\algtext*{EndSwitch}%
\algtext*{EndCase}%

\usepackage[capitalise]{cleveref}

\RequirePackage{pgfplots}
\pgfplotsset{compat=newest}
\usetikzlibrary{plotmarks}
\usetikzlibrary{arrows.meta}
\usetikzlibrary{shapes}
\usepgfplotslibrary{units}
\usepgfplotslibrary{patchplots}
\usepackage{tikzscale}
\usepackage{tikz}
\usetikzlibrary{arrows,positioning,shapes.geometric,arrows.meta,decorations.markings} 

\tikzstyle{basic box}[black] = [shape=rectangle, align=center, draw=#1, rounded corners] 
\tikzstyle{basic box b}[red] = [shape=rectangle, align=center, draw=#1, rounded corners]
\tikzstyle{state} = [circle, draw, very thick]

\tikzstyle{basic box b}[red] = [shape=rectangle, align=center, draw=#1, rounded corners]

\tikzset{
    >=stealth',
    punkt/.style={
           rectangle,
           rounded corners,
           draw=black, very thick,
           text width=6.5em,
           minimum height=2em,
           text centered},
    pil/.style={
           ->,
           thick,
           shorten <=2pt,
           shorten >=2pt,},
    flash/.style args={#1:#2}{postaction=decorate,decoration={name=markings,
    mark=at position #1 with {%
    \draw[fill=#2, line width=.75\pgflinewidth, line cap=round, line join=round]
         (+\pgflinewidth,+7\pgflinewidth)   -- ++ ( left:+2\pgflinewidth) 
      -- (+-4\pgflinewidth,+-\pgflinewidth) -- ++ (right:+5\pgflinewidth)
      -- (+-\pgflinewidth,+-7\pgflinewidth) -- ++ (right:+2\pgflinewidth)
      -- (+4\pgflinewidth,\pgflinewidth)    -- ++ (left:+5\pgflinewidth)
      -- cycle;}}}
}

\begin{document}
%
\title{Web-Based Platform for Evaluation of Resilient and Transactive Smart-Grids}


%
\author{\IEEEauthorblockN{Himanshu Neema, Harsh Vardhan, Carlos Barreto, and Xenofon Koutsoukos}
\IEEEauthorblockA{Institute for Software Integrated Systems, Vanderbilt University, Nashville, TN, USA
}}


\maketitle

\begin{abstract}
Today's smart-grids have seen a clear rise in new ways of energy generation, transmission, and storage. This has not only introduced a huge degree of variability, but also a continual shift away from traditionally centralized generation and storage to distributed energy resources (DERs). In addition, the distributed sensors, energy generators and storage devices, and networking have led to a huge increase in attack vectors that make the grid vulnerable to a variety of attacks. The interconnection between computational and physical components through a largely open, IP-based communication network enables an attacker to cause physical damage through remote cyber-attacks or attack on software-controlled grid operations via physical- or cyber-attacks. Transactive Energy (TE) is an emerging approach for managing increasing DERs in the smart-grids through economic and control techniques. Transactive Smart-Grids use the TE approach to improve grid reliability and efficiency. However, skepticism remains in their full-scale viability for ensuring grid reliability. In addition, different TE approaches, in specific situations, can lead to very different outcomes in grid operations. In this paper, we present a comprehensive web-based platform for evaluating resilience of smart-grids against a variety of cyber- and physical-attacks and evaluating impact of various TE approaches on grid performance. We also provide several case-studies demonstrating evaluation of TE approaches as well as grid resilience against cyber and physical attacks.

\end{abstract}


%
\IEEEpeerreviewmaketitle

\section{Introduction}
\label{sec:introduction}

The power-grids of today are transforming dramatically in a number of ways to become much more digitally controlled as well as intelligently operated \cite{santacana2010getting} \cite{conti2014annual}. Several new features have emerged that make a power-grid more 'smart'. The key aspect being increased participation by consumers in the demand response. The increased digital connectivity and access to sophisticated meters has enabled the power consumers to make more informed decisions about how much and when to consume power. In addition, increased availability of DERs (e.g., solar panels, wind turbines, and storage batteries) has enabled consumers to send excess power back to the grid. These consumers are referred to as `\textit{prosumers}' as they both produce and consume power. Thus, the model of previously largely centralized electrical grid is morphing heavily into a rather distributed grid with smart meters, and one that must be managed for demand response in a dynamic manner.

\begin{figure}[htb] 
\centering
\includegraphics[width=0.98\columnwidth,]{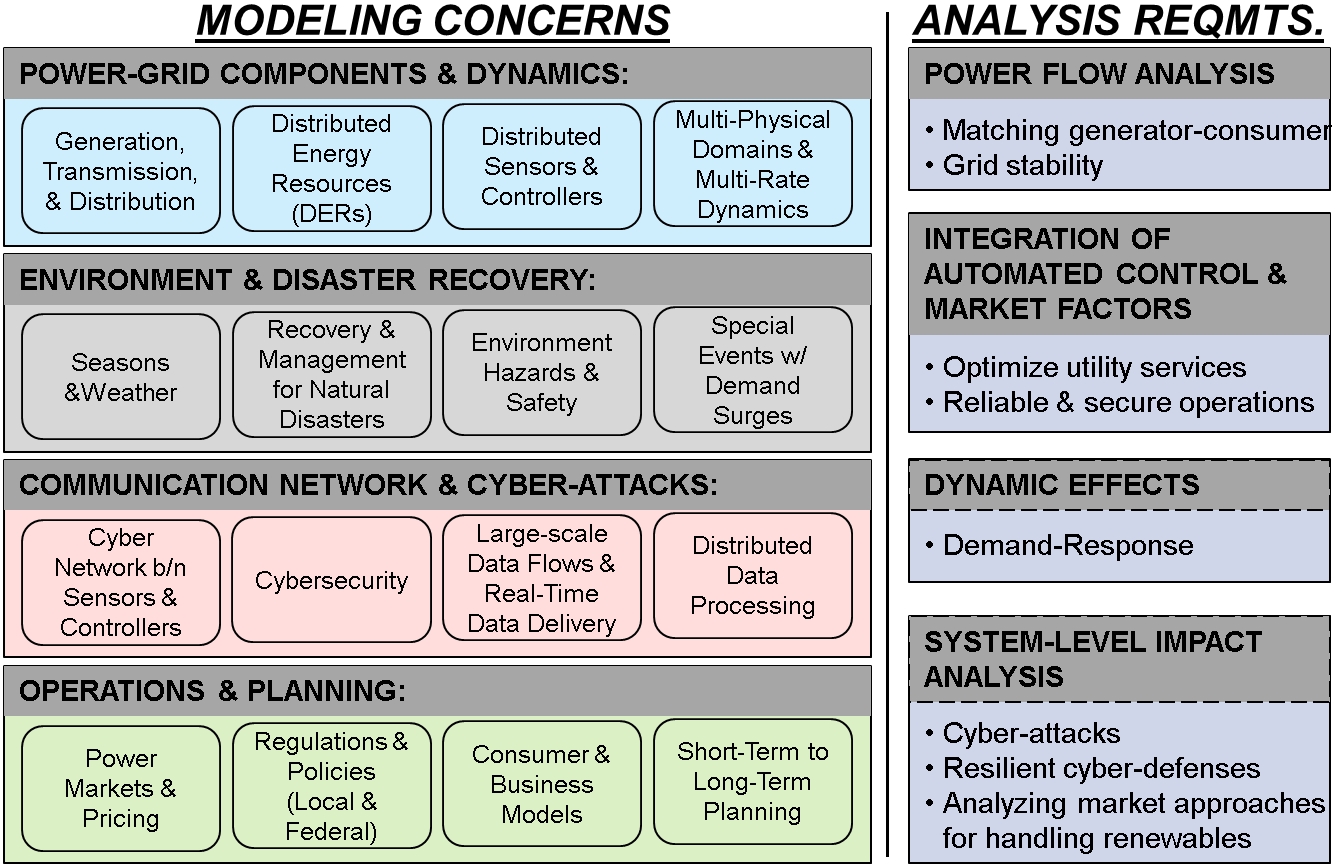}
\caption{Smart-Grid Modeling \& Evaluation Requirements}
\label{fig:modeling_concerns}
\end{figure}

\cref{fig:modeling_concerns} depicts a large number of concerns that must be addressed for a holistic evaluation of smart-grids as well as several higher-level objectives for their end-to-end analysis.

The increase in DERs and sophisticated control methods has enabled multiple power markets and mirco-grids to co-exist. This makes power-grids even more dynamic and complex to manage. \textit{Resilient power-grids} have enough redundancies, security mechanisms, and diversity built into its components and systems such that the overall grid operations are not excessively impacted even in the presence of cyber- and/or physical disruptions. \textit{Reliable power-grids} are not only resilient, but also provide an assurance of the availability, quality, and security of the power supply. Managing resilience and reliability of smart-grids in a way that addresses all of the concerns depicted in \cref{fig:modeling_concerns} is highly challenging. Yet the DERs and digital connectivity among them enables one to operate grid more efficiently by dynamically managing the demand response through economic and digital control techniques such as Transactive Energy (TE) \cite{gridwise2015gridwise}. However, analyzing different TE approaches and evaluating grid for resilience against attacks requires a comprehensive platform that supports end-to-end analysis of a variety of grid configurations.

A significant concern due to increased DERs and digital control is a huge increase in grid's attack surface. Grids are more vulnerable to a variety of attacks, both cyber and physical. Grid's physical components and the digital monitoring and control equipment are interconnected using largely Internet Protocol (IP) based communication networks. This can enable adversaries to deploy sophisticated cyber-attacks to disrupt the critical grid operations and even cause blackouts. For example, the cyber-attack deployed on Ukraine power-grid caused many substations to fail leading many communities without power \cite{liang20172015}. Adversaries could also deploy \textit{integrity} attacks that modify network packets in a subtle manner such that the attack becomes apparent only after its cumulative effect over time. In addition, faults can be introduced to cause spikes in demand response, and as these components exchange sensor data and control inputs, attacks causing faults in one component can result in \textit{cascading faults} throughout the grid.


Continuous monitoring using edge devices and smart meters can help with detecting attacks on the grid. However, operating the grids that deploy large number of DERs is still highly challenging \cite{camacho2011control}. Evaluation of grid operations with DERs and transactive energy systems, along with impact of cyber- and physical-attacks, require a comprehensive platform where different combinations of such scenarios can be analyzed for a variety of grid configurations.

This paper describes a novel web-based platform that allows evaluating smart-grids for resilience against cyber-physical attacks and for their behavior when different transactive energy approaches are used. The platform provides a highly configurable and extensible web-based metamodeling environment to model power-grids, market behavior, and various attack configurations. This platform also provides a \textit{simulation backend} for power-grid simulation (using GridLAB-D \cite{chassin2008gridlab}) of the modeled grid configurations and experiments. Section~\ref{sec:architecture} presents the overall architecture of the platform including the workflow for using it for smart-grid evaluations, as well as how the entire system is implemented. In Sections~\ref{sec:resilience_eval} and~\ref{sec:te_eval}, we respectively describe how power-grids should be evaluated against cyber-physical attacks and for transactive energy approaches. A few relevant use-cases are demonstrated with experiment results in Section~\ref{sec:experiments}. Finally, Section~\ref{sec:conclusions} concludes the paper and highlights directions for future work.
\section{Platform Architecture}
\label{sec:architecture}

\begin{figure}[htb] 
\centering
\includegraphics[width=0.8\columnwidth,]{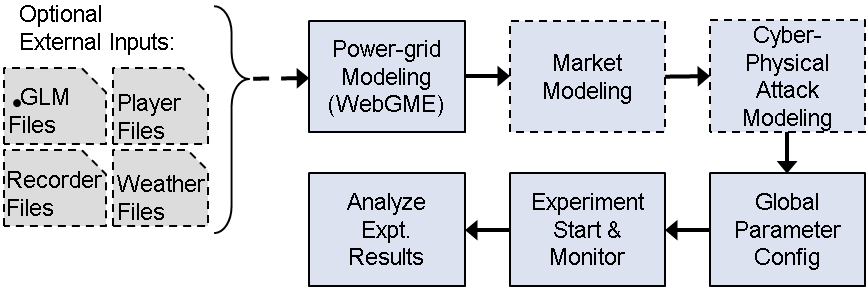}
\caption{Modeling \& Simulation Workflow}
\label{fig:studio_workflow}
\end{figure}

In this research, we aim to provide a web-based platform \cite{design-studio-url} for modeling and simulating power-grids that supports evaluation of grid operations for different transactive energy (TE) approaches and evaluating grid's resilience against a variety of cyber-physical attacks. It is accessible via a web-browser and does not require any software installations. \cref{fig:studio_workflow} shows the modeling and simulation workflow for using the platform. The optional elements are shown by dashed boxes. The user starts with modeling the grid, which can also be input optionally as an external input as a \textit{GridLAB-D model} file (.GLM). Other configurations for time-series parameter values can be input as \textit{player} files. Statistics to collect from simulation can be specified in \textit{recorder} files. \textit{Weather} information - relevant to many grid modules like solar panels - can also be provided. The grid model can directly be simulated in the \textit{backend}. Several \textit{global} parameters are also supported such as the physical time being simulated (e.g., to match against weather data). Researchers can also experiment with different market and cyber-physical attack models in the grid simulation. Upon simulation completion, the results including recorder statistics are given to the user. Continues feedback is also presented to monitor the simulation progress.

\begin{figure}[htb] 
\centering
\includegraphics[width=0.85\columnwidth,]{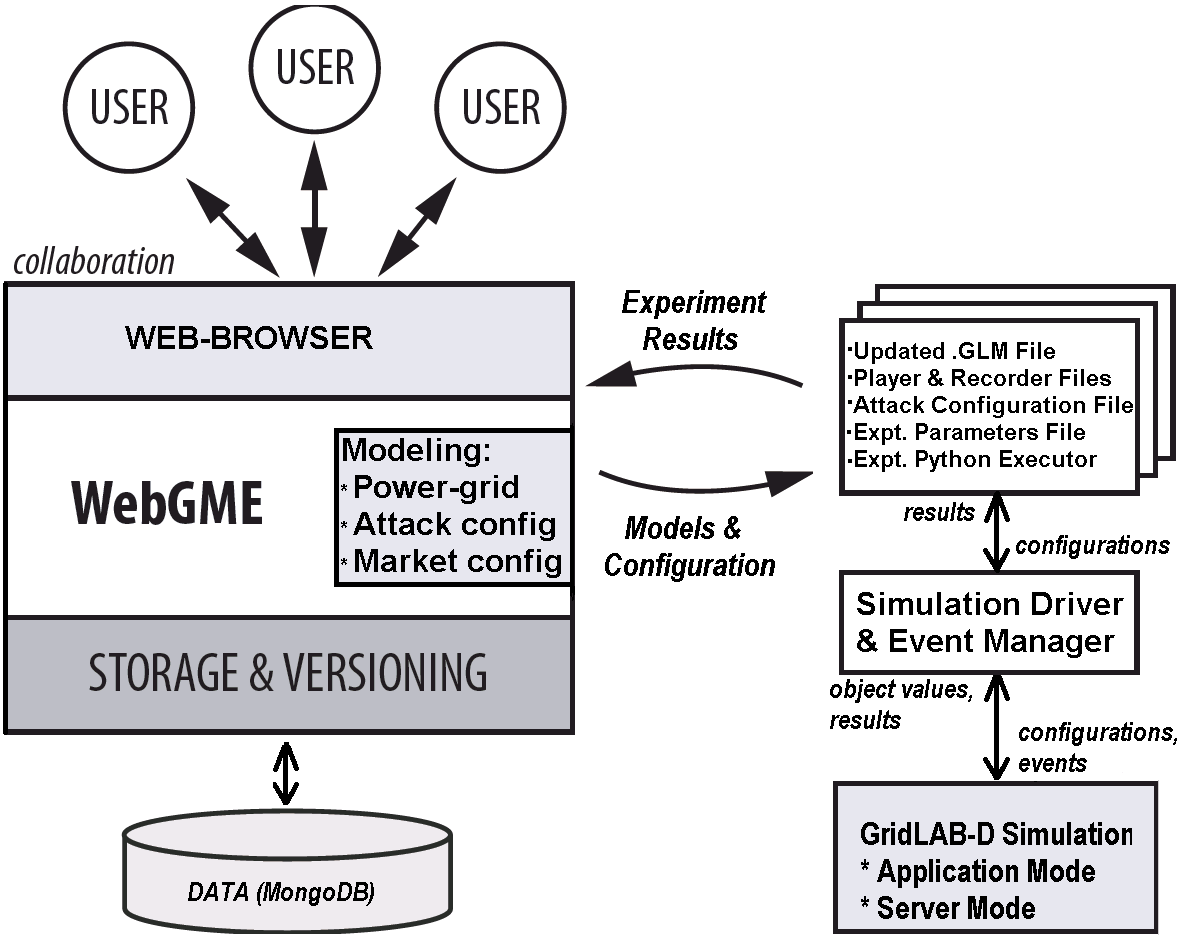}
\caption{Implementation Architecture}
\label{fig:studio_architecture}
\end{figure}

\subsection{Overview}

As shown in \cref{fig:studio_architecture}, the platform utilizes a web-based metamodeling framework, called WebGME \cite{maroti2014online}, to provide a graphical modeling language for designing power-grids based on concepts from the GridLAB-D simulator. The language also allows modeling different market approaches and attack configurations. When GridLAB-D adds or modifies its concepts, those can be automatically imported in the language. A model interpreter can automatically synthesize executable code and associated configurations for setting up and running corresponding power-grid simulations. Experiment results are also provided to the user using the same interface. WebGME also permits multiple users to collaborate on the model simultaneously by maintaining change history and versions through a MongoDB database. Our platform is hosted on the CPS-VO portal \cite{cps-vo} that enables creating users and user-groups, managing authentication and authorization, and providing secure and private collaboration among group users.

\begin{figure}[htb] 
\centering
\includegraphics[width=0.98\columnwidth,height=5.5cm]{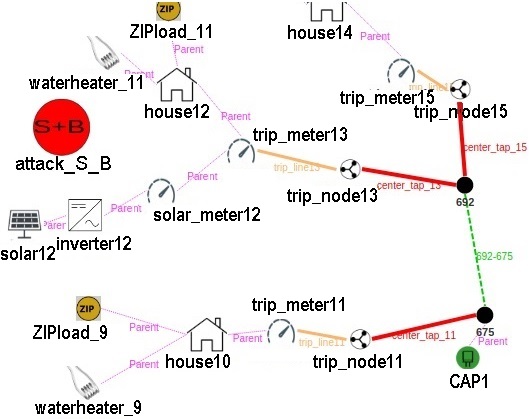}
\caption{Part of Sample Grid Model with Attack Config.}
\label{fig:sample_grid}
\end{figure}

\cref{fig:sample_grid} shows part of a sample grid modeled in our platform. The full grid model consists of two \textit{nodes} connected through \textit{underground line} (green dashed line). Each \textit{node} is connected to primary side of center-taped \textit{transformer}. The secondary side of \textit{transformer} is connected with \textit{triplex node}, which itself is connected to \textit{triplex meters} using \textit{triplex overhead lines}. There is one \textit{triplex meter} for each \textit{house}, which provides net-metering. Each \textit{house} is connected through a \textit{zip-load} and a \textit{water-heater}. One of the \textit{net-meter} (triplex meter) is connected to a \textit{solar-inverter} pair through another meter which measures the power generated by \textit{solar-panel}. The \textit{solar-panel} generates power based on climate data (as TMY3 file from NOAA), power rating, and other parameters. The grid is simulated using initial parametric values and using \textit{forward-back sweep} method. Component 'attack\_S\_B' (in red) is an \textit{attack configuration} for both seller and buyer prices. The platform provides a library of many such components.

\subsection{Simulation Backend}

The backend of our platform consists of a \textit{Simulation Driver \& Event Manager} module (Section~\ref{sec:sim_driver_evt_mgr}) and a configurable GridLAB-D simulation infrastructure in a cloud environment. We chose GridLAB-D power-grid simulator as it is open-source, calculates power-flow variable values before every simulation step, and supports modeling of end-use load models that incorporate weather and market behaviors. This enables modeling of distribution systems and evaluation of DERs and TE approaches. However, precise description of events must still be supported efficiently when a large number of power-grid components are used. An additional challenge is that GridLAB-D restricts the parameters to be linear combinations of variables.
Our platform overcomes these challenges by extending high-level modules in GridLAB-D for supporting power-grid simulations with capability to model market variations as well as attacks that are \textit{specific} (e.g., on a single component) or \textit{generic} (i.e., affecting multiple components).

\subsection{Simulation Driver \& Event Manager}
\label{sec:sim_driver_evt_mgr}

As shown in \cref{fig:studio_architecture}, the WebGME generated artifacts are sent to a \textit{Simulation Driver (SM) \& Event Manager (EM)} module. \textit{SM} executes and controls the simulation. The simulation is executed in \textit{server mode} if attacks or market approaches are configured because that requires adding new events during run-time. Algorithm ~\ref{algo:event_manager} describes the behavior of \textit{EM}. \textit{EM} processes the schedules specified in WebGME to create  corresponding events-list for grid objects. The simulation loop incorporates the actions with intervening pauses during which object values are updated.

\begin{algorithm} 
\caption{Event manager}
\label{algo:event_manager} 
\begin{algorithmic}[1] 
    \Require List of schedules, start time $t_0$, and stop time $t_f$.
    \State Pause the simulation at $t_0$.
    \State Add future events from schedules in an ordered list.
    \State $t \leftarrow t_0$
    \While{ $t\leq t_f$ }
        \State Wait for a pause in the simulation.
        \State Update the simulation time $t$.
        \For {Each event occurring at time $t$}
	  \State Execute the event.
	  \State Update the list of future events. 
        \EndFor
	\State Continue the simulations.
    \EndWhile
\end{algorithmic}
\end{algorithm}

\subsection{Modeling Physical Attacks}

We leverage the \emph{reliability} module from GridLAB-D to model physical attacks that disconnect segments of the grid.
The reliability module 
allows the analysis of the system's state before and after a fault, but ignores
the transients between these states.
Moreover, it only handles faults in lines, fuses, and switches, causing topology changes (e.g., disconnections).

We simulate attacks by creating events that change the connectivity of the system's conductors
(e.g., current flow can be interrupted by changing status of lines to `\textit{\textsc{open}}').
Disconnection of specific switches and fuses
is also supported.

\begin{figure}
\centering
\resizebox{.6\columnwidth}{!}{
\begin{tikzpicture}[edge from parent path={(\tikzparentnode.west) .. controls +(-.4, 0) and +(.4, 0)
.. (\tikzchildnode.east)}]
 
  \tikzstyle{edge from parent}=[black, thick, draw]
  \tikzstyle{level 1} = [sibling distance=20pt, level distance=130pt]

\node[basic box, very thick] (market) at (0,0) {Market};

\node[basic box, very thick, right of=market, node distance=80pt] (seller) {Seller};

\node[basic box, very thick, left of=market, node distance=80pt] (buyer) {Buyer};

\draw[punkt, <->, color=black] (market) --   node[below, font=\small,] {}  (seller);

\draw[line width=1.4pt, line cap=rect]
  (market) edge[flash=.5:red]  (seller);

\draw[punkt, <->, color=black] (market) --   node[below, font=\small,] {}  (buyer);

\path[line width=1.4pt, line cap=rect]
  (market) edge[flash=.5:red] (buyer) ;

  \path  (market) -- (seller) node[below, pos=0.5](att1){}  ;
  \path  (market) -- (buyer) node[below, pos=0.5](att2){}  ;

  \node[below of=market, node distance=40pt] (attacker) {Attack};
  
  \draw [punkt, thick] (attacker.east) edge[bend right=35, ->] (att1.south);
  \draw [punkt, thick] (attacker.west) edge[bend left=35, ->] (att2.south);
  
  
\end{tikzpicture}
}
\caption{Exploiting Market Infrastructure for Attacks}
\label{fig:normal_market}
\end{figure}
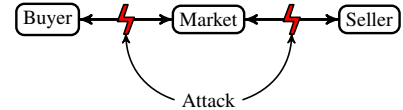

\begin{figure}
\centering
\resizebox{.7\columnwidth}{!}{
\begin{tikzpicture}[edge from parent path={(\tikzparentnode.west) .. controls +(-.4, 0) and +(.4, 0)
.. (\tikzchildnode.east)}]
 
  \tikzstyle{edge from parent}=[black, thick, draw]
  \tikzstyle{level 1} = [sibling distance=20pt, level distance=130pt]

\node[very thick, basic box] (market) at (0,0) {Main\\Market}; 
\node[basic box, below of=market, node distance=60pt] (aux_market) {Auxiliary\\Market};

\node[very thick, basic box, right of=market, node distance=70pt] (seller) {Seller};
\node[very thick, basic box, left of=market, node distance=110pt] (buyer) {Buyer};

\node[basic box, left of=market, node distance=60pt] (bidder) {Auxiliary\\bidder};
\node[basic box, below of=seller, node distance=60pt] (bidder2) {Auxiliary\\bidder};

\draw[<->, color=black, punkt] (aux_market) -|  node[below, font=\small,] {}  (buyer);
\draw[<->, color=black, punkt] (market) --  node[below, font=\small,] {}  (seller);

\draw[<-, color=black] (aux_market) --   node[below, font=\small,] {}  (bidder2);
\draw[<-, color=black] (bidder2) --   node[below, font=\small,] {}  (seller);

\draw[<->, color=black] (aux_market) -|   node[below, font=\small,] {}  (bidder);

\draw[<-, color=black] (market) --   node[below, font=\small,] {}  (bidder);

\end{tikzpicture}
}
\caption{False Data Injection via Auxiliary Market and Bidder}
\label{fig:implementation}
\end{figure}
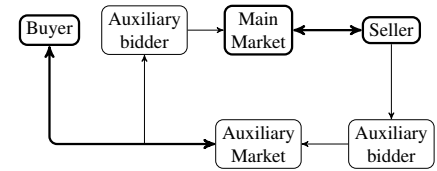

\subsection{GridLAB-D Extensions}

\cref{fig:normal_market} shows the general structure of a market that accepts bids from both buyers and sellers. 
Attacks on this market type of system
usually pursues two objectives, namely damage the system or profit from the attack.
GridLAB-D models represent mostly physical connections among components, except communications for market interactions.
A communication network among market participants is not supported. Instead, the components communicate through internal functions.

The market's communication structure makes it difficult to implement the false data injection attacks.
The bids sent by controllers to the \textit{Main Market (MM)} cannot be directly modified, but we can estimate them. Secondly,
the controllers observe the market's clearing prices, which invalidates the reported fake prices. 
We handle these by introducing \textit{Auxiliary Market (AM)} (see \cref{fig:implementation}). Here,
the buyer bids directly in \textit{AM} and makes decisions based on the prices in \textit{AM}.
We estimate the bids because precide bids are not observable.
Thus, we implement an \textit{Auxiliary Bidder (AB)} that sends the estimated bids to \textit{MM}.
\textit{AB} also precisely replicates seller's bids in \textit{AM} - possible because in our model seller's bids are constant.

Without external interventions, the models in \cref{fig:normal_market,fig:implementation} are equivalent. Both markets will reach the same equilibrium if \textit{ABs} make the same bids as the buyer and seller.\footnote{In practice, the equilibrium of both the main and the auxiliary markets have minor differences because of errors and delays calculating the bids.} However, the modification shown in \cref{fig:implementation} enables one to modify the bids by targeting the \textit{ABs}.

\section{Evaluating Grid Resilience against Attacks}
\label{sec:resilience_eval}

Smart-grids utilize a large number of components which exchange data through largely IP-based communication networks for relaying data. The increased digital connectivity and DERs have made grids highly vulnerable to many different types of cyber-physical attacks. The cyber-attacks could even be carried out in a subtle manner such that they go unnoticed for long periods of time or exploit connectivity among devices to reach remote systems through multi-hop network propagation. Even the physical-attacks on devices can lead to transients in the grid causing faults to propagate to remote regions that were the real intended targets of the attacker.

The attacks mainly serve two purposes for the attackers, viz. derive profit by manipulating the market's prices or damage the system by creating imbalances in the system, which translate into frequency instability.
For example, 
in the aurora attack study, Idaho National Lab used fast opening and closure of the circuit breakers of a generator that desynchronized it and even damaged it permanently.
Other attacks against the transmission system could even disconnect lines, transformers, and generators, thus changing the topology of the system.

Depending on how the attack directly impacts the grid operations, they can be classified as \textit{specific} (i.e., on a single component such as solar-panel or a load) or \textit{generic} (i.e., affecting multiple components such as varying the market behavior).
In our platform, we support modeling of these types of attacks and executing the power-grid simulation to determine grid's performance amidst these attacks.

\section{Evaluating Transactive Energy Approaches}
\label{sec:te_eval}

Transactive energy (TE) \cite{gridwise2015gridwise} is a distributed management approach that 
expands electricity markets into the retail domain.
The TE paradigm relies on economic and control mechanisms to balance supply and demand dynamically.
Some widely analyzed TE price schemes are \emph{time of use} (TOU) rates, \emph{critical peak pricing} (CPP), and \emph{real time pricing} \cite{siano2014demand}.
However, unlike some price mechanisms, the TE can implement
a market to allow transactions 
among the system's elements.
However, TE creates uncertainities about grid operations due to the use of DERs, widely varying consumer behaviors, and increased risks associated with the large number of connected and heterogeneous components
\cite{weerakkody2019challenges, barreto2018impact}.
Both industry and academia have focused their efforts \cite{hammerstrom2008pacific, ohio2016gridsmart, melton2015pacific} on evaluating the impact of TE on grid operations.
In addition, NIST has organized a large program specifically to address challenges with TE \cite{nist-te-challenge}.

In our platform, we utilize the modules available in GridLAB-D to support evaluation of TE approaches. For example, using the transactive controller modules - which manage appliances and submit market bids based on both current prices and system's state \cite{fuller2011analysis} - we support modeling different market mechanisms  \cite{broeer2014modeling} such as \textit{buyers-only}, \textit{sellers-only}, and \textit{auctions}. Our platform supports not only the evaluation of different TE approaches, but also allows incorporating cyber-physical attacks on the grid at the same time. Taken together, the platform can be a powerful tool for analyzing a variety of power-grid scenarios with many different types of market and attack models.


\section{Experiment Results}
\label{sec:experiments}




\subsection{Power System Model}
In these experiments, we make a detailed simulation of an electric distribution system using GridLAB-D and the prototypical distribution feeder models provided by the 
 Pacific Northwest National Laboratory (PNNL) \cite{osti_1040684}.
The distribution models capture fundamental characteristics of distribution utilities from the U.S. 
In this case, we use the 
prototypical feeder \emph{R1-12.47-3}, which represents a moderately populated area.
Furthermore, we added representative residential loads to the distribution model using the script in \cite{pop_script}.

In summary, our distribution model has $109$ commercial and residential loads, which in turn incorporate appliances such as heating, ventilation, and air conditioning (HVAC) systems, water heaters, and pool pumps.
GridLAB-D allows us to model the response of the loads to weather and market's prices, giving realism to the simulations. 
In particular, we emulate the temperature in Nashville, TN, 
during summer time. 
On the other hand, we assume that the system has $50$ generators, that can output a maximum of 2 MW.

\subsection{Attacks Utilized}

In our attack scenarios, we illustrate attacks that
directly manipulate control commands or inject false data in the system.
%
In \textit{Attack Scenario 1},
the adversary exploits the market infrastructure to influence users and induce a peak in demand. 
Thus, the attacker has to manipulate the market in such a way that the users concentrate their demand  at a particular time.
The success of this attack depends on both the attack resources and the response of users to the market's equilibrium. 

In \textit{Attack Scenario 2}, 
the adversary attempts to manipulate the market's equilibrium to profit (e.g., with higher market clearing prices).
In this case, the success depends solely on the attacker's capacity to manipulate the market's equilibrium, in other words, in its resources to implement the attack.

\subsection{Attack Scenario 1: Create Peaks in Demand}

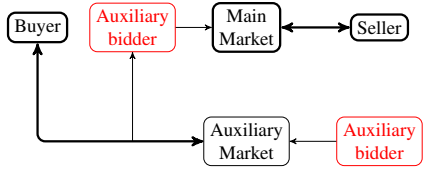
\begin{figure}
\centering
\resizebox{.7\columnwidth}{!}{
\begin{tikzpicture}[edge from parent path={(\tikzparentnode.west) .. controls +(-.4, 0) and +(.4, 0)
.. (\tikzchildnode.east)}]
 
  \tikzstyle{edge from parent}=[black, thick, draw]
  \tikzstyle{level 1} = [sibling distance=20pt, level distance=130pt]

\node[very thick, basic box] (market) at (0,0) {Main\\Market}; 
\node[basic box, below of=market, node distance=60pt] (aux_market) {Auxiliary\\Market};

\node[very thick, basic box, right of=market, node distance=70pt] (seller) {Seller};
\node[very thick, basic box, left of=market, node distance=110pt] (buyer) {Buyer};

\node[basic box, left of=market, node distance=60pt, color=red] (bidder) {Auxiliary\\bidder};
\node[basic box, below of=seller, node distance=60pt, color=red] (bidder2) {Auxiliary\\bidder};

\draw[<->, color=black, punkt] (aux_market) -|  node[below, font=\small,] {}  (buyer);
\draw[<->, color=black, punkt] (market) --  node[below, font=\small,] {}  (seller);

\draw[<-, color=black] (aux_market) --   node[below, font=\small,] {}  (bidder2);

\draw[<->, color=black] (aux_market) -|   node[below, font=\small,] {}  (bidder);

\draw[<-, color=black] (market) --   node[below, font=\small,] {}  (bidder);

\end{tikzpicture}
}
\caption{Market Price Manipulation via Auxiliary Bidders}
\label{fig:attack_1}
\end{figure}

This attack attempts to create peaks in demand by modifying the prices that the  controllers (buyers) observe. 
Due to the system's restrictions, the adversary cannot change directly the prices that the controllers observe. Instead, it can modify the bids submitted by sellers to the auxiliary market (see \cref{fig:attack_1}).
In particular, an adversary with enough resources can decide the future market's price 
by setting all the bids with the same price.
Thus, in the worst case, the adversary has the power to (indirectly) set the prices.

Although manipulating the market's price is essential, the adversary cannot succeed unless the users  (or more precisely, the transactive controllers) increase their demand at a particular time.
Since the power system has mechanisms to deal with uncertainties in demand and operational failures,  the adversary must achieve a fast and large imbalance in the system.
Hence, the coordination of users plays a critical role.

In this scenario, an adversary targets HVAC with its attack. 
These controllers choose the appliances set points (which in turn determine the energy consumption) based on the system's state (internal and outdoor temperature) and both the current and previous prices.
For instance, the transactive controllers 
can turn off (on) the HVAC when it observes high (low) prices to reduce the cost of energy.
The precise decisions follow the preferences of users (e.g., their tolerance to temperature changes as a function of monetary compensations).

In this case the adversary leverages the behavior of the transactive controllers to create peaks in demand.
The attack first increases the price, which forces the HVAC to use less energy at the expenses of allowing higher indoor temperatures.
Later, when the indoor temperature is high enough, the attacker suddenly reduces the price. This encourages higher energy usage to reduce the indoor temperature. However, this signal has a coordinating effect that raises the energy demand almost simultaneously, resulting in a demand peak.

\subsubsection{Experiments}

\begin{figure}[htb] 
\centering
\includegraphics[width=0.9\columnwidth,]{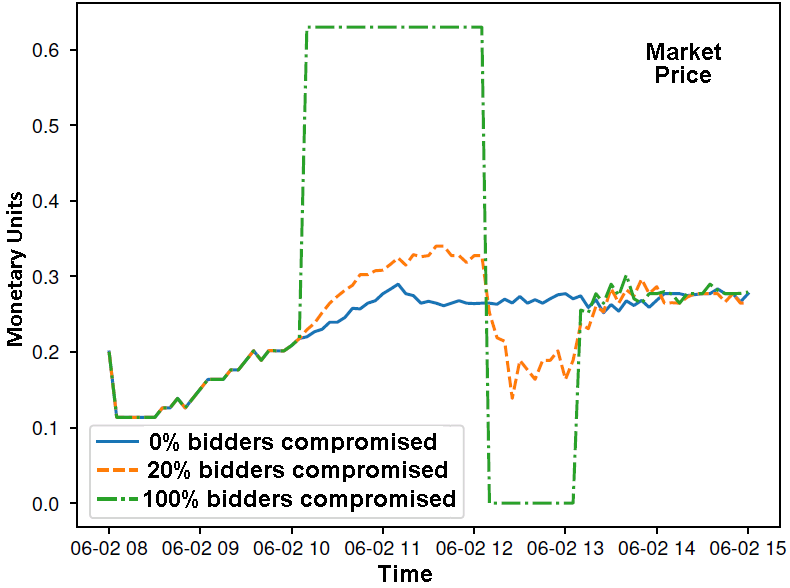}
\caption{Market Price Changes by Modified Seller Bids}
\label{fig:price1}
\end{figure}

\begin{figure}[htb] 
\centering
\includegraphics[width=0.9\columnwidth,]{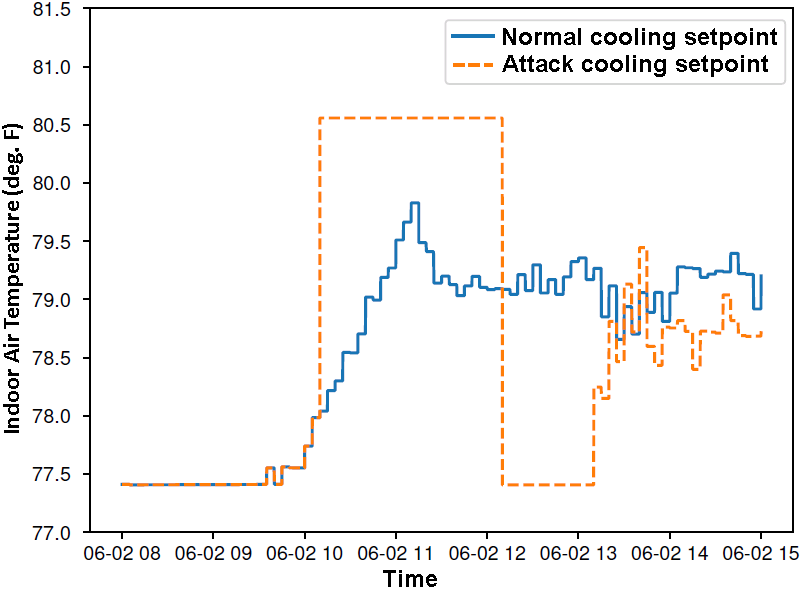}
\caption{Indoor Temperature Setpoint of HVAC System}
\label{fig:temp}
\end{figure}

\cref{fig:price1} illustrates the impact of attacks with different resources (fraction of bidders compromised) that causes a demand peak at 12 pm.
The attack starts at 10 am, when the adversary modifies the seller's bids to the maximum price accepted by the market ($0.63$). The high price signal induces demand reduction because the controllers choose a higher cooling set point (see \cref{fig:temp}).   Despite of the sudden price change, the controllers do not react simultaneously; hence, this first stage doesn't harm the system.

\begin{figure}[htb] 
\centering
\includegraphics[width=0.9\columnwidth,]{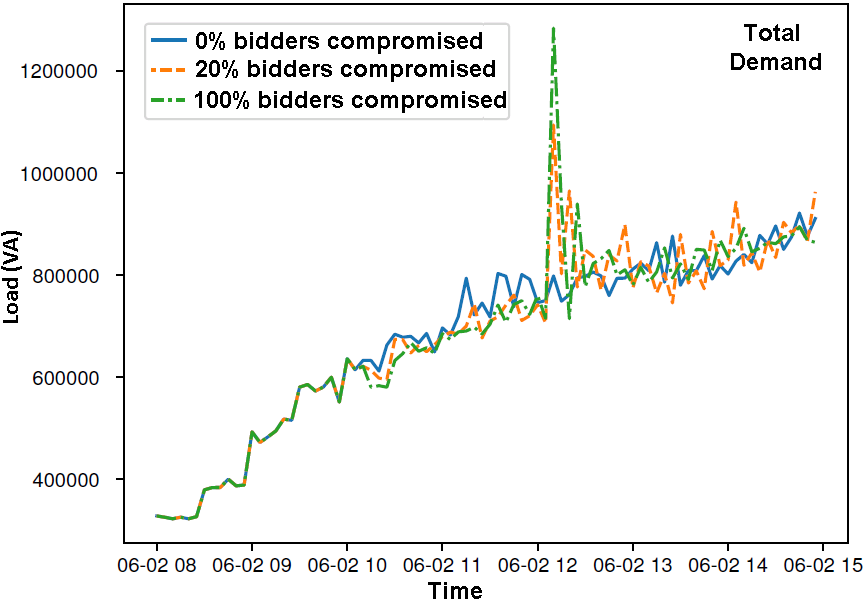}
\caption{Demand Peak Created in Attack Scenario 1}
\label{fig:demand1}
\end{figure}

The adversary then lowers the prices suddenly at 12 pm, which changes the cooling set points of HVAC systems (see \cref{fig:temp}). 
By this time, most of the houses had high temperatures. Consequently, the HVACs immediately turn on their cooling systems, creating a demand peak.
 \cref{fig:demand1} shows
that peak demand can be created even when attacker compromises only $20\%$ of bidders. despite the lower impact on the market's price.

An attack's timing is crucial for its success. In particular, the price change can induce a peak because most of the houses reached high temperatures, which created a coordinating effect in the demand. Thus, the adversary needs continuous access to the cooling system and enough time to execute the attack.




\subsection{Attack Scenario 2: Modifying the Controllers' Bids}

This attack seeks to change the bids of the buyers to raise the prices and benefit the sellers. 
The adversary compromises the bids of  buyers (controllers) and sends the following price:
\begin{equation}
 \hat p_c = p_c + \lambda p_m,
\end{equation}
where $p_c$ is the price submitted by the controller, $p_m$ is the market price, $\lambda$ is the control parameter, and $\lambda \geq 0$ represents the intensity of the attack.
When $\lambda = 0$ the attack doesn't change the bids; however, $\lambda>0$ raises the bid for energy, which increases the market's clearing price.
The adversary may select a small value of $\lambda$ to prevent attack discovery.

\subsubsection{Experiments}

\begin{figure}[htb] 
\centering
\includegraphics[width=0.9\columnwidth,]{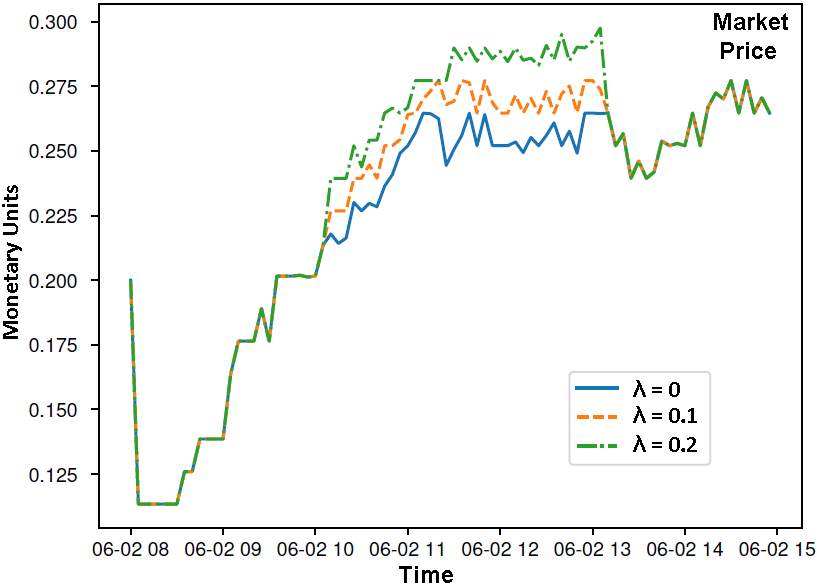}
\caption{Deviation in Market Price in Attack Scenario 2}
\label{fig:price2}
\end{figure}

\cref{fig:price2} show the market prices in the second attack scenario (the attack runs from 10am till 13pm). The deviation from the normal price occurs due to the attack that raises the prices, and depends on the parameter $\lambda$.


\section{Conclusions \& Future Work}
\label{sec:conclusions}

Power-grids are a complex system involving many different components. Recent increase in DERs have put tight operational constraints for balancing power demand and generation. In addition, the increase in digital connectivity among grid components and controllers has significantly increased grid's vulnerability to cyber- and physical-attacks.

Transactive Energy (TE) methods can mitigate some demand response issues and improve resource utilization and operational efficiency. For safe and reliable grid operations to ensure availability, security, and reliability of power supply, the grid must incorporate security mechanisms and redundant and diverse components. These are complex considerations that must be effectively used based on comprehensive evaluations.

In this paper, we described a web-based platform that enables researchers to evaluate such complex considerations simply via a web-browser. We presented the platform's architecture and discussed how it can be used to evaluate resilience of smart-grids against a variety of cyber- and physical-attacks and to model various TE approaches and evaluate them for grid performance. We also provided several case-studies with experiment results that demonstrate the platform's capabilities for conducting such modeling and simulation experiments.

In the future, we plan to extend the platform to allow specification of a design of experiments (DOE) for different combinations of topological and parametric variations and to automatically execute simulations for all such combinations. We also plan to extend and further parameterize the current library of reusable models for grid components, markets, and attack configurations. Better visualization of experiment results both during and after simulation is also proposed.
We are also currently applying the platform's tools and techniques for higher-level evaluations such as societal implications of the ongoing increase in renewable energy implementations.
\section*{Acknowledgments}
\label{sec:acks}



This work is supported in part by the NSF FORCES program under grant CNS-1238959, by the NSF PIRE and CPS program under award \#1521617, and by NIST under awards \#70NANB18H269 and \#70NANB17H266.

\balance
\begin{spacing}{0.88}
\bibliographystyle{IEEEtran}
\bibliography{references,references_b}

\begin{thebibliography}{10}
\providecommand{\url}[1]{#1}
\csname url@samestyle\endcsname
\providecommand{\newblock}{\relax}
\providecommand{\bibinfo}[2]{#2}
\providecommand{\BIBentrySTDinterwordspacing}{\spaceskip=0pt\relax}
\providecommand{\BIBentryALTinterwordstretchfactor}{4}
\providecommand{\BIBentryALTinterwordspacing}{\spaceskip=\fontdimen2\font plus
\BIBentryALTinterwordstretchfactor\fontdimen3\font minus
  \fontdimen4\font\relax}
\providecommand{\BIBforeignlanguage}[2]{{%
\expandafter\ifx\csname l@#1\endcsname\relax
\typeout{** WARNING: IEEEtran.bst: No hyphenation pattern has been}%
\typeout{** loaded for the language `#1'. Using the pattern for}%
\typeout{** the default language instead.}%
\else
\language=\csname l@#1\endcsname
\fi
#2}}
\providecommand{\BIBdecl}{\relax}
\BIBdecl

\bibitem{santacana2010getting}
E.~Santacana, G.~Rackliffe, L.~Tang, and X.~Feng, ``Getting smart,'' \emph{IEEE
  Power and Energy Magazine}, vol.~8, no.~2, pp. 41--48, 2010.

\bibitem{conti2014annual}
J.~J. Conti, P.~D. Holtberg, J.~A. Beamon, A.~M. Schaal, J.~Ayoub, and J.~T.
  Turnure, ``Annual energy outlook 2014,'' \emph{US Energy Information
  Administration}, 2014.

\bibitem{gridwise2015gridwise}
{GridWise Architecture Council}, ``Gridwise transactive energy framework:
  Version 1.0,'' \emph{Pacific Northwest National Laboratory, PNNL-22946 Ver
  1.0}, 2015.

\bibitem{liang20172015}
G.~Liang, S.~R. Weller, J.~Zhao, F.~Luo, and Z.~Y. Dong, ``The 2015 ukraine
  blackout: Implications for false data injection attacks,'' \emph{IEEE
  Transactions on Power Systems}, vol.~32, no.~4, pp. 3317--3318, 2017.

\bibitem{camacho2011control}
E.~F. Camacho, T.~Samad, M.~Garcia-Sanz, and I.~Hiskens, ``Control for
  renewable energy and smart grids,'' \emph{The Impact of Control Technology,
  Control Systems Society}, pp. 69--88, 2011.

\bibitem{chassin2008gridlab}
D.~P. Chassin, K.~Schneider, and C.~Gerkensmeyer, ``Gridlab-d: An open-source
  power systems modeling and simulation environment,'' in \emph{Transmission
  and distribution conference and exposition, 2008. t\&d. IEEE/PES}.\hskip 1em
  plus 0.5em minus 0.4em\relax IEEE, 2008, pp. 1--5.

\bibitem{design-studio-url}
\BIBentryALTinterwordspacing
``{Power-grid analysis framework},'' Feb. 2019. [Online]. Available:
  \url{https://cps-vo.org/group/gridlabd}
\BIBentrySTDinterwordspacing

\bibitem{maroti2014online}
M.~Mar{\'o}ti, R.~Keresk{\'e}nyi, T.~Kecsk{\'e}s, P.~V{\"o}lgyesi, and
  A.~L{\'e}deczi, ``Online collaborative environment for designing complex
  computational systems,'' \emph{Procedia Computer Science}, vol.~29, pp.
  2432--2441, 2014.

\bibitem{cps-vo}
\BIBentryALTinterwordspacing
``{Cyber-Physical Systems Virtual Organization (CPS-VO)},'' Feb. 2018.
  [Online]. Available: \url{https://cps-vo.org}
\BIBentrySTDinterwordspacing

\bibitem{siano2014demand}
P.~Siano, ``Demand response and smart grids—a survey,'' \emph{Renewable and
  sustainable energy reviews}, vol.~30, pp. 461--478, 2014.

\bibitem{weerakkody2019challenges}
S.~Weerakkody and B.~Sinopoli, ``Challenges and opportunities: Cyber-physical
  security in the smart grid,'' in \emph{Smart Grid Control}.\hskip 1em plus
  0.5em minus 0.4em\relax Springer, 2019, pp. 257--273.

\bibitem{barreto2018impact}
C.~Barreto and A.~Cardenas, ``Impact of the market infrastructure on the
  security of smart grids,'' \emph{IEEE Transactions on Industrial
  Informatics}, pp. 1--1, 2018.

\bibitem{hammerstrom2008pacific}
D.~J. Hammerstrom, R.~Ambrosio, T.~A. Carlon, J.~G. DeSteese, G.~R. Horst,
  R.~Kajfasz, L.~L. Kiesling, P.~Michie, R.~G. Pratt, M.~Yao \emph{et~al.},
  ``Pacific northwest gridwise™ testbed demonstration projects; part i.
  olympic peninsula project,'' Pacific Northwest National Lab.(PNNL), Richland,
  WA (United States), Tech. Rep., 2008.

\bibitem{ohio2016gridsmart}
A.~Ohio, ``gridsmart (sm) demonstration project,'' 2016.

\bibitem{melton2015pacific}
R.~Melton, ``Pacific northwest smart grid demonstration project technology
  performance report volume 1: Technology performance,'' Pacific Northwest
  National Lab.(PNNL), Richland, WA (United States), Tech. Rep., 2015.

\bibitem{nist-te-challenge}
\BIBentryALTinterwordspacing
``{NIST Transactive Energy Challenge},'' Feb. 2019. [Online]. Available:
  \url{https://pages.nist.gov/TEChalleng}
\BIBentrySTDinterwordspacing

\bibitem{fuller2011analysis}
J.~C. Fuller, K.~P. Schneider, and D.~Chassin, ``Analysis of residential demand
  response and double-auction markets,'' in \emph{Power and Energy Society
  General Meeting, 2011 IEEE}.\hskip 1em plus 0.5em minus 0.4em\relax IEEE,
  2011, pp. 1--7.

\bibitem{broeer2014modeling}
T.~Broeer, J.~Fuller, F.~Tuffner, D.~Chassin, and N.~Djilali, ``Modeling
  framework and validation of a smart grid and demand response system for wind
  power integration,'' \emph{Applied Energy}, vol. 113, pp. 199--207, 2014.

\bibitem{osti_1040684}
K.~P. Schneider, Y.~Chen, D.~P. Chassin, R.~G. Pratt, D.~W. Engel, and S.~E.
  Thompson, ``Modern grid initiative distribution taxonomy final report,''
  Pacific Northwest National Laboratory, Tech. Rep., 2008.

\bibitem{pop_script}
\url{https://github.com/gridlab-d/Taxonomy_Feeders/tree/master/PopulationScript},
  2015, accessed: January 12, 2019.

\end{thebibliography}
\end{spacing}
%

\end{document}